\documentclass[preprint,superscriptaddress,floatfix]{revtex4-1}
\usepackage[]{graphicx}
\usepackage{amsmath}
\usepackage{amsfonts}
\usepackage{amssymb}
\usepackage{tabularx}

\begin{document}
\title{Electronic Structure of Magnetic Semiconductor CdCr$_2$Te$_4$: A Possible Spin-Dependent Symmetry Filter}
\author{H. Sims}
\affiliation{Center for Materials for Information Technology and Department of Physics,\\
University of Alabama, Tuscaloosa AL 35487-0209}
\author{K. Ramasamy}
\affiliation{Center for Materials for Information Technology and Department of Chemistry,\\
University of Alabama, Tuscaloosa AL 35487-0209}
\author{W. H. Butler}
\affiliation{Center for Materials for Information Technology and Department of Physics,\\
University of Alabama, Tuscaloosa AL 35487-0209}
\author{A. Gupta}
\affiliation{Center for Materials for Information Technology and Department of Chemistry,\\
University of Alabama, Tuscaloosa AL 35487-0209}
\begin{abstract}
We present a theoretical investigation of the electronic and magnetic structure of spinel CdCr$_2$Te$_4$ using density functional theory, its extensions via onsite Hubbard $U$ interactions, and a screened-hybrid-functional exchange potential. We find that the ground state is semiconducting within the latter approach, and within this magnetic-semiconducting system we compute the complex band structure, finding a slowly decaying evanescent $\tilde{\Delta}_1$ state possibly suitable for realizing a spin-dependent symmetry filter effect.
\end{abstract}
\maketitle

Chromium chalcospinels have been heavily studied for their interesting electronic, optical, and magnetic properties for decades.\cite{baltzer,Menyuk,awang,Ramasamy2012,Ramasamy2011,RamSims2013,kanomata,suzuzhang,gap1,gap2,gap3,Antonov1999,Zhang2012} In particular, the (Cu,Cd)Cr$_2$X$_4$ (X = S, Se, Te) series show a wide range of electronic properties. Wang {\it et al.}\cite{awang} studied Cd$_x$Cu$_{1-x}$Cr$_2$S$_4$ and Cd$_x$Cu$_{1-x}$Cr$_2$Se$_4$ within the generalized gradient approximation, reporting that half metals could be created from the metallic (CuCr$_2$(S,Se)$_4$) and semiconducting (CdCr$_2$(S,Se)$_4$) base systems at x = 0.5. Among the tellurides, CuCr$_2$Te$_4$ has been reported to be a ferromagnetic metal\cite{Ramasamy2012,kanomata,suzuzhang} with a Curie temperature near 300 K. We are not aware of any experimental or theoretical studies of CdCr$_2$Te$_4$, but the properties of CdCr$_2$S$_4$ and CdCr$_2$Se$_4$ suggest that CdCr$_2$Te$_4$ should also be semiconducting, likely with a rather small gap due to the periodic trends within the chalcogen group.

Looking beyond the current and upcoming generations of spintronic devices, efficient spin injection into semiconductors may provide a path to achieving low-power technologies for processing, nonvolatile storage and memory, and other novel applications. Magnetic semiconductors and insulators are of great interest for such devices, as their spin-dependent barrier height should lead to highly spin-polarized tunneling currents without the need for a half-metallic electrode. The spin-filter effect is based on this concept; a properly-designed magnetic tunnel barrier (or double barrier) could in principle provide the large on/off ratio needed for spin logic. The europium chalcogenides are the prototypical spin filters,\cite{santos} but their low $T_c \sim$ 70 K, 17 K, and 5 K for EuO, EuS, and EuSe, respectively, limit their usefulness in real applications (the Curie temperatures of the Cu-based chromium chalcospinels are generally above 300 K). Other applications envision using a magnetic semiconductor as a spin aligner, a layer that polarizes incoming current and then injects spins into a nonmagnetic semiconducting material, such as a GaAs/AlGaAs light-emitting diode.\cite{Fiederling1999}

In the following, we will present calculations that suggest that CdCr$_2$Te$_4$ is not only a magnetic semiconductor but in fact a potential spin-dependent symmetry filter. We predict that CdCr$_2$Te$_4$ has majority- and minority-channel band gaps of 0.49 eV (indirect) and 0.62 eV (direct), respectively. Further, we compute the complex band structure in both channels at $\mathbf{k}_\parallel = 0$, finding that the minority gap is spanned by a slowly-decaying evanescent state possessing $\tilde{\Delta}_1$ symmetry.

In our investigation, we employ calculations within density functional theory\cite{dft} (DFT)---using the generalized-gradient approximation of Perdew, Burke, and Ernzerhof (PBE),\cite{pbe} ``+$U$'' corrections through the Dudarev formulation of LDA+$U$,\cite{ldau,Dudarev} and the Heyd-Scuseria-Ernzerhof (HSE06) hybrid functional method,\cite{hse06} which has been shown to improve on the accuracy of LSDA or PBE band gaps, particularly in semiconductors and small-gap insulators.\cite{vasphse} All calculations were completed using the {\sc vasp} code\cite{vasp} using the projector augmented-wave (PAW)\cite{paw} pseudopotentials of Kresse and Joubert.\cite{pp,moredetails} In the PBE+U calculation, we apply $U_{eff} = 2.5 eV$ to the Cr 3$d$ orbitals. This value is typical for octahedrally-coordinated magnetic Cr ions.\cite{Sims2010,Yaresko2008,Gupta2013} We find that the qualitative electronic structure (i.e.~metallic, half-metallic, or semiconducting) of this material is insensitive to changes in the interaction parameter. All calculations were performed using the primitive 14-atom supercell (space group $Fd\mathit{\bar{3}}m$) of the normal spinel structure. In all cases, we consider the PBE-relaxed structure with lattice constant $a =$ 11.491 \AA\ and internal parameter $x \approx$ 0.264 (equivalent to $u =$ 0.389 in the more commonly reported setting). The real and complex band structures were computed using a Wannier basis obtained via the {\sc Wannier90} package.\cite{w90}

\begin{table}[h!tb]
\caption{Calculated Electronic and Magnetic Structure of other chalcospinels within HSE06. The lattice parameters and internal parameters were calculated within PBE and are in good agreement with experiment. The magnetic moments and band gaps (if any) are calculated within HSE06. Experimental data are those quoted in Refs.~\onlinecite{awang} and \onlinecite{Antonov1999} unless otherwise noted. The HSE06 band gap for CdCr$_2$S$_4$ is spin dependent, but we only report the smallest gap here for comparison with experiment.}\label{tab:otherchalc}
\begin{ruledtabular}
\begin{tabularx}{\columnwidth}{lrlrc}
 & \multicolumn{2}{c}{Mag. Mom.} & \multicolumn{2}{c}{Gap (eV)} \\
 & \multicolumn{2}{c}{($\mu_B$/f.u.)} & \multicolumn{2}{c}{} \\
& HSE06 & Expt. & HSE06 & Expt. \\
\cline{2-3} \cline{4-5}
CuCr$_2$S$_4$ & 5.08 & 5.20 & --- & --- \\
CuCr$_2$Te$_4$ & 5.52 & 5.42 & --- & --- \\
CdCr$_2$S$_4$ & 6.00 & 6.00 & 2.2 & 1.99--2.61\footnote{Refs.~\onlinecite{gap1,gap2,gap3}}\\
\end{tabularx}
\end{ruledtabular}
\end{table}

In the following, we will show that the HSE06 screened hybrid functional method is alone among the methods we employ in predicting a semiconducting ground state for CdCr$_2$Te$_4$. It is therefore necessary to determine whether this result is trustworthy. As there are no experimental results against which to compare, we attempted to determine whether the screened hybrid-functional method could accurately describe the electronic structure of other materials in the (Cu,Cd)Cr$_2$(S,Te)$_4$ family. We summarize the results in Table~\ref{tab:otherchalc}. In each of these calculations, we use the PBE-relaxed structure. First, we see that HSE06 correctly predicts the presence of the band gap in CdCr$_2$S$_4$ while correctly identifying CuCr$_2$S$_4$ and CuCr$_2$Te$_4$ as metals. The magnitude of the gap in CdCr$_2$S$_4$ reported in the literature varies between 2 and 2.6 eV,\cite{gap1,gap2,gap3} with our calculated gaps fitting into that range. In contrast, PBE and PBE+U (using the same $U_{eff}$ = 2.5 eV) predict band gaps of $\sim 1$ eV and 1.3 eV, respectively.

It has been reported that HSE06 can badly overestimate the magnetic moment in bcc Fe\cite{Jang,Sims2012} and possibly other magnetic metals. However, we do not observe any such problem in these results. Comparing with Refs.~\onlinecite{awang} and \onlinecite{Antonov1999}, we find that our moments are all in reasonably good agreement with experiment and with past computations. This is likely due to the more localized nature of the Cr magnetic moments. Considering both the level of agreement between our calculations and past experimental and computational reports and the indications in the PBE and PBE+$U$ electronic structure of an incipient semiconducting state, we propose that we can draw conclusions from our HSE06 results with some confidence.

\begin{figure}[h!tb]
\includegraphics[]{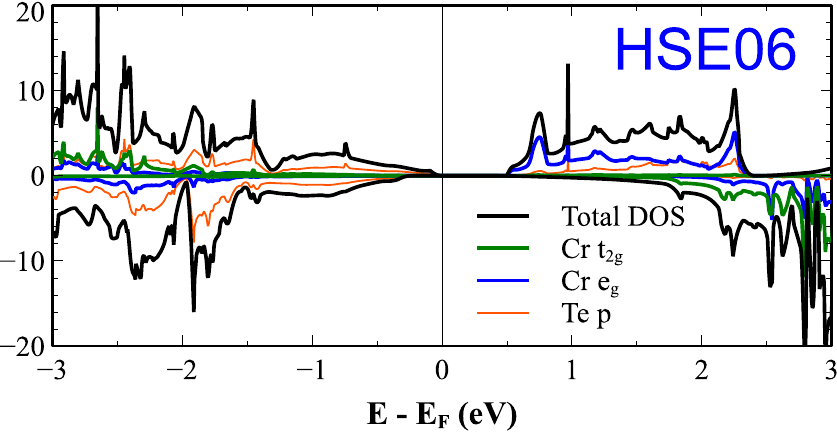}
\caption[]{Density of states for CdCr$_2$Te$_4$ for the HSE06 method. The site- and $l$-decomposed partial densities of states for the Cr 3$d$ and Te 5$p$ states are provided in each plot. Note that only HSE06 is capable of completely opening the gap in both spin channels.}\label{fig:alldos}
\end{figure}

We find that within PBE and PBE+$U$, CdCr$_2$Te$_4$ is nearly half-metallic, with a minority valence band composed almost entirely of Te 5$p$ states.\cite{suppl} Incorporating a portion of the Fock exchange potential using the HSE06 method yields a slightly different picture (Figure~\ref{fig:alldos}). As in PBE, the top of the valence band has predominantly Te $p$ character, but the Fock exchange pushes all other occupied states deeper, with the Cr $t_{2g}$ now lying between -6.3 and -1.2 eV. A gap is created in the majority channel, and the gap in the minority states is widened further, resulting in a ferromagnetic semiconductor with spin-dependent band gaps---about 0.49 eV and 0.62 eV in the majority and minority channels, respectively. The differing orbital character of the majority and minority band edges gives rise to this spin-dependence, also seen in other magnetic insulating spinels (see Ref.~\onlinecite{Sun2012}, \onlinecite{Hollinsworth2013}, and \onlinecite{Caffrey2013}, in which similar behavior is reported in NiFe$_2$O$_4$ and CoFe$_2$O$_4$). This should be a general property of insulating or semiconducting spinels (and perhaps other ternary transition-metal oxides and chalcogenides) in which the valence states of one (possibly magnetic) transition metal species sit well below the valence band and the other species is magnetic. Such a property may allow for the individual tuning of majority and minority band gaps via, for example, the electronegativity of the anion.

\begin{figure}[h!tb]
\includegraphics[]{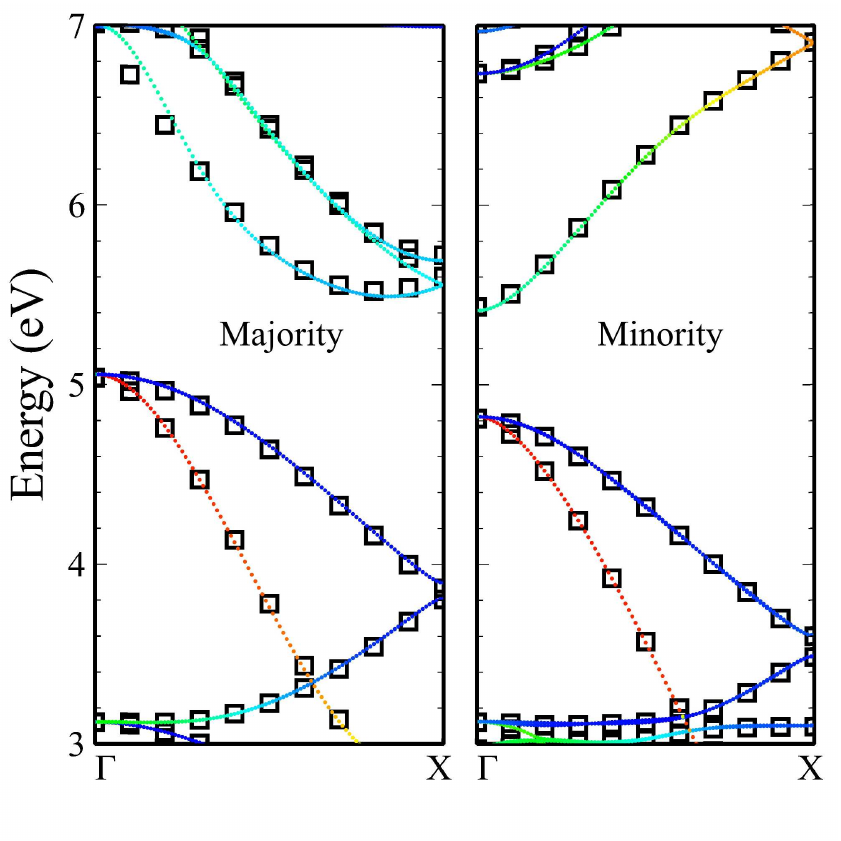}
\caption{Band structure for CdCr$_2$Te$_4$ within HSE06, with the majority spin channel shown in the left panel and the minority channel on the right. We note that the majority gap is indirect, while the minority gap is direct at $\Gamma$. The open symbols represent bands computed directly within HSE06, while the dotted lines are those obtained with a Wannier Hamiltonian. The color of the latter is indicative of the $p_z$ or $d_{xy}$ character of the band (with warmer colors indicating more orbital character). Near $\Gamma$, the bottom of the minority conduction band also has Cd $s$ character. We note that the majority CBM has significant $d_{3z^2-r^2}$ character as well.}\label{fig:hsebands}
\end{figure}

The HSE06 band structure along the line $\Gamma$ -- X $\equiv$ 2$\pi$/$a$ (0 0 1) can be found in Figure~\ref{fig:hsebands}, with the majority bands on the left and the minority bands on the right. The open circles are the bands calculated directly within the electronic structure method, while the dotted lines are computed from a Wannier basis containing only Cr $d$ and Te $p$ states. In the minority channel, one can see that the bands are in near-perfect agreement, particularly near the Fermi energy. The small error near $\Gamma$ in the minority channel is due to our exclusion of the Cd $s$ states. The error is somewhat greater in the majority channel, although once again the accuracy is best near the band edges. We note that the HSE calculation on which the Wannier bands are based uses a finer $k$ mesh, and so it is possible that not all of the error lies in the Wannier bands. The agreement of the bands calculated in the two methods is of special importance here as we will use the Wannier Hamiltonian to compute the complex band structure.

Examination of the bands  shows that the majority gap is predicted to be indirect between $k_z$ = 0 and $k_z \approx 0.84 \frac{2\pi}{a}$, while the larger minority gap is direct at $\Gamma$. The valence bands are composed almost exclusively of Te $p$ states in both channels and are therefore identical apart from a small shift. The majority CBM appears to consist of a Cr $d_{3z^2-r^2}$ band, while the minority CBM comprises Cd $s$ and Cr $d_{xy}$ states, with the latter orbital character becoming stronger as $k_z$ approaches the zone boundary. In Figure~\ref{fig:hsebands}, the Wannier bands are colored according to their $p_z$ or $d_{xy}$ character.

\begin{figure}[h!tb]
\includegraphics[]{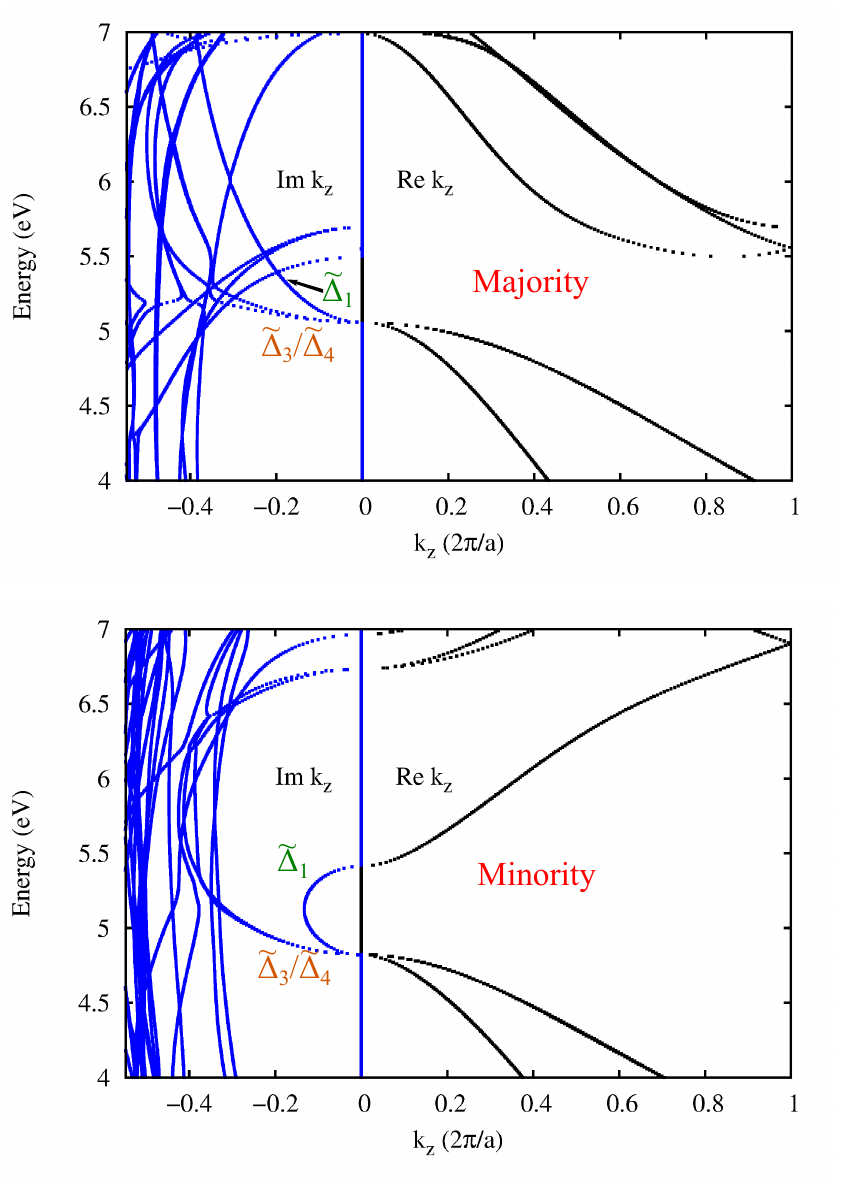}
\caption{Complex band structure for CdCr$_2$Te$_4$ computed from a Wannier Hamiltonian obtained from the HSE06 calculations for majority (top) and minority (bottom) channels. Purely real bands are shown to the right with positive $k$, while the imaginary parts of the complex bands are on the left of the plots with negative $\kappa$. Note the lone evanescent state in the minority gap connecting the top of the valence band with the bottom of the conduction band. The next higher conductions bands are also connected to the $p_x$ and $p_y$ states through a complex $\tilde{\Delta}_3$/$\tilde{\Delta}_4$ state.}\label{fig:cbands}
\end{figure}

In the prototypical Fe/MgO symmetry filter, one finds that bands with $\Delta_1$ symmetry are much more easily propagated through the MgO tunnel barrier than the $\Delta_2$, $\Delta_{2^\prime}$, or $\Delta_5$ bands, and that these $\Delta_1$ bands only cross the Fermi level in one spin channel in Fe, giving rise to a highly spin-polarized current.\cite{butlermgo,femgoreview} The $C_{2v}$ point group of spinel oxides yields bands with $\tilde{\Delta}_1$, $\tilde{\Delta}_2$, $\tilde{\Delta}_3$, and $\tilde{\Delta}_4$ symmetry. As explained in Ref.~\onlinecite{Zhang2012}, the $\tilde{\Delta}_1$ bands have both the $s$, $p_z$, $d_{3z^2-r^2}$ character of the square-symmetric $\Delta_1$ bands and $\sim d_{xy}$ character of the $\Delta_2$/$\Delta_{2^\prime}$ bands. We therefore would like to determine whether the band at the minority CBM and the dispersive $p_z$ band at the valence edge are part of the same complex $\tilde{\Delta}_1$ band and are thus suitable for symmetry filtering. We accomplish this by computing the complex band structure according to the method of Chang and Schulman\cite{CScbands} within the same Wannier basis employed to compute the real band structure. The results can be seen in Figure~\ref{fig:cbands}. Evanescent states decay with characteristic length inversely proportional to the imaginary wave vector $\kappa$, and so we are primarily concerned with those states in which $\kappa$ remains small throughout the gap. We note that, although both channels contain states with rapidly increasing $\kappa$, there is a single state in the minority gap that joins the VBM and CBM with $\kappa \lesssim 0.13 \frac{2\pi}{a} \approx 0.07$ \AA$^{-1}$. This slowly decaying evanescent state completes the $\tilde{\Delta}_1$ band and gives rise to the possibility of symmetry filtering in the minority spin channel. Of course, as noted in Ref.~\onlinecite{Zhang2012}, $\tilde{\Delta}_1$ also matches $\Delta_2$/$\Delta_{2^\prime}$ bands, and so further investigation would be necessary to determine the extent of the filtering.

Following the argument of Ref.~\onlinecite{butlermgo}, we can estimate the decay of the $\tilde{\Delta}_1$ band as exp(-2$\kappa\Delta z$), where $\Delta z = c$ is the $z$ distance between the CdCr$_2$Te$_4$ (conventional) unit cells. As noted above, $\kappa \lesssim 0.07$ \AA$^{-1}$, so $\kappa \Delta z = \sqrt{-(k_{z,\tilde{\Delta}_1}\Delta z)^2} \approx 0.817$ at the middle of the gap. In comparison, $\kappa \Delta z \approx 1.47$ in MgO.\cite{butlermgo} One can see in Figure~\ref{fig:cbands} that there is another semicircular evanescent state joining $\tilde{\Delta}_3$/$\tilde{\Delta}_4$ bands, with a maximum $\kappa$ of $\sim$ -0.4 $2\pi/a$ $\approx$ 0.22 \AA$^{-1}$ and $\kappa \Delta z \lesssim$ 2.51.

In the vicinity of the gap, it is possible (for $\Delta_1$-like bands) to express $k^2$ as
\begin{equation}
\frac{1}{k^2(E)} = \frac{\hbar^2}{2m_{v}^{*}(E - E_v)} + \frac{\hbar^2}{2m_{c}^{*}(E_c - E)}\label{eq:k2}
\end{equation} 
where $E_v$ and $E_c$ are the top of the valence band and the bottom of the conduction band, respectively, and $m_{v}^{*}$ and $m_{c}^{*}$ are the corresponding effective masses (of the $\tilde{\Delta}_1$ band). Fitting this expression to our computed complex bands within the gap, we can estimate $m_{c}^{*} = m_{v}^{*} \approx 0.15 m_e$. This is in reasonable agreement with the value obtained by fitting only the real-$k$ valence and conduction band edges ($\approx 0.178 m_e$) using the first and second terms in \eqref{eq:k2}, respectively. The fits accurately reproduce the complex bands near the edges of the gap and within the gap,where the $\Delta_1$ character is most prominent. We can use the second term of the free-electron like Eq.~\eqref{eq:k2} for the majority spin, allowing us to compute $m_{c}^{*} \approx 0.53 m_e$.

To summarize, we have employed PBE, PBE+U, and the HSE06 screened hybrid functional method in our investigation of the relatively unexplored spinel CdCr$_2$Te$_4$. All three predict a ferromagnetic ground state with a moment of $\sim 3 \mu_B$ per Cr ion. Within the former two methods, CdCr$_2$Te$_4$ is described as a highly spin-polarized zero-gap semiconductor/near half-metal. The HSE06 method, which gives a better account of the electronic structure of the related CdCr$_2$S$_4$ system and accurately describes other chromium chalcospinels, increases the gap to form a true magnetic semiconductor with majority and minority band gaps of 0.49 eV and 0.62 eV, respectively. In this semiconducting state, we estimate the effective mass of the conduction electrons to be $m^{*}_{\uparrow} = 0.53 m_e$ and $m^{*}_{\downarrow}$ = 0.15 $m_e$ -- 0.18 $m_e$.
Further, we find evidence of a $\tilde{\Delta}_1$ complex band spanning the band gap in the minority channel. The direct gap, effective masses, and possible symmetry-filter effect in the minority channel may lead to electron currents strongly polarized with minority spin, despite the smaller gap in the majority channel.

Only experimental and theoretical investigations into realistic heterostructures containing CdCr$_2$Te$_4$ can truly predict the prospects for application. Nevertheless, the spin-dependence of the symmetry filtering, band-edge transitions, band gaps, and conduction electron effective masses provides fertile ground for optimization and design of future spin-based technologies.

The authors wish to acknowledge funding from an NSF-NRI supplement as part of NSF MRSEC Grant No. DMR-0213985  and NSF CHE-1012850. H. S. and W.H.B. also wish to acknowledge conversations with J. Zhang and X.-G. Zhang. All calculations were performed on the University of Alabama MINT High Performance Cluster.


\begin{thebibliography}{00}
\bibitem{moredetails} We use a plane-wave cutoff of 274.3 eV (20.16 Ry). A $16\times16\times16$ $\Gamma$-centered Monkhorst-Pack mesh of $k$-points was necessary to achieve convergence, with a finer $20\times20\times20$ mesh used for the PBE and PBE+$U$ density of states calculations. For the computation of the short-range Fock exchange potential in the HSE06 method, the $k$ mesh was reduced to $3\times3\times3$ (but remained at $12\times12\times12$ for other portions of said calculation).
\end{thebibliography}
\end{document}